# Fair Recommendation by Geometric Interpretation and Analysis of Matrix Factorization

Hao Wang
haow85@live.com
Ratidar.com
Beijing, China

**ABSTRACT**

Matrix factorization-based recommender system is in effect an angle preserving dimensionality reduction technique. Since the frequency of items follows power-law distribution, most vectors in the original dimension of user feature vectors and item feature vectors lie on the same hyperplane. However, it is very difficult to reconstruct the embeddings in the original dimension analytically, so we reformulate the original angle preserving dimensionality reduction problem into a distance preserving dimensionality reduction problem. We show that the geometric shape of input data of recommender system in its original higher dimension are distributed on co-centric circles with interesting properties, and design a paraboloid-based matrix factorization named ParaMat to solve the recommendation problem. In the experiment section, we compare our algorithm with 8 other algorithms and prove our new method is the most fair algorithm compared with modern day recommender systems such as ZeroMat and DotMat Hybrid.

## 1. INTRODUCTION

Today, every big internet company is into recommender systems. Recommender systems can boost traffic volume and increase sales revenues by a large margin (30%-40% for all sales on Amazon) while saving a huge amount of marketing budget. Although recommender systems suffered from a major setback in the first half of 2010's when most companies agreed unanimously that recommender system could not serve as a stand-alone product, in a short period after the rampant spreading of pessimistic opinions, companies such as TikTok and Kuai Shou emerged as major players on global market as stand-alone recommender system products.

Researchers invented recommender system a couple of decades ago, and the trickling stream of the AI technology becomes torrents today. The basic idea behind recommender system is to use big data analytical mechanisms to analyze historic information of users and items to recommend new items to users based on computed preference scores. For example, if we know Alice loves reading *Su Tung-Po* , *The Three Kingdoms* and many other Chinese Classics, we could recommend other Chinese Classics which she had not read for her. This procedure is much more effective to help Alice find her new books than search engines with which Alice needs to find what she likes by herself. The example we just illustrated is the primitive form of a recommender system technology named *content-based recommendation*. It is just one of the sub-fields of many recommender system research areas.

Since the year of 2016, prestigious research venues such as RecSys [1][2] has witnessed the rise of deep neural network models. With more and more companies and schools taking up the course, the technical models of recommender systems are becoming more and more complex and individualistic. Since deep neural networks can represent any function, it enables researchers to choose the best model from a much larger candidate pool than in the age of shallow models. However, due to the number of choices in parameter tweaking, public knowledge of deep neural network becomes more and more individualistic – usually only a small hand of researchers know the reasons and principles behind the neural models that they produce.

Although earlier models in the age of shallow models seem out-of-dated , they are still widely used in various companies. One of the major reasons of their longevity is their simplicity and interpretability. Due to the same reasons, we choose matrix factorization [3] as our main research target in this paper. We provide a geometric interpretation and analysis of the classic matrix factorization algorithm, and based on our observations and analysis, we propose a new algorithm named ParaMat that is much more interpretable , and at the same time very accurate when compared with other models.

## 2. RELATED WORK

Recommender system remains a popular research field irrespective of the up-and-down in the investment in the field. As one of the most successful recommendation paradigms, matrix factorization is provided with a probabilistic framework as the foundation in 2007 [4]. The probabilistic framework is named Probabilistic Matrix Factorization. Based on modification of this framework, ZeroMat [5] and DotMat [6] are proposed to solve the cold-start and sparsity problem without input data. The algorithms could predict user preferences fairly accurately with no historic information. The social implications of the 2 algorithms are astonishing - human cultures are locked into a predictable state only after a couple of years of evolution.

Matrix factorization can also be used to solve the context-aware recommendation (CARS) problem [7] [8]. In 2021, a new CARS solution named MatMat [9] is introduced. Instead of scalar fitting, MatMat uses matrix factorization by matrix fitting to incorporate contextual information into the system. A practical example of MatMat named MovieMat [10] which solves CARS problem for movie recommendation is proposed in the same year. The algorithm incorporates no more than 6 contextual information fields in the algorithm and achieves better results than classic models.

Fairness is a hot research topic in recent years. Google comes up with an algorithm named Focused Learning [11] in the year of 2017 which penalizes the matrix factorization loss function with a fairness metric. More researchers spend time and energy on fair Learning to Rank approaches [12] [13] [14] and publish extensively at top conferences such as SIGIR and KDD. However, due to its simplicity, matrix factorization still remains a good benchmark for fairness ideas. Zipf Matrix Factorization [15] is introduced in 2021 with the introduction of a fairness metric named Degree of Matthew Effect as a side product. MatRec [16] and KL-Mat [17] are also examples of fair recommender algorithms based on matrix factorization framework. In 2022, Wang [18] proposed a set of fairness metrics using extreme value theory.

## 3. GEOMETRIC ANALYSIS OF MATRIX FACTORIZATION

One of the most popular definition of the loss function of matrix factorization is as follows :

$$L = \sum_{i=1}^{n}\sum_{j=1}^{m} \left(R_{i,j} - U_i^T \cdot V_j\right)^2 \tag{1}$$

In this formula, $R_{i,j}$ represents the rating value that the i-th user gives on the j-th item. $U_i$ is the user feature vector, and $V_j$ is the item feature vector. If $U_i$ and $V_j$ are not normalized in the real world applications, we usually encounter gradient explosion problems, so a more practical loss function of matrix factorization is actually defined as follows :

$$L = \sum_{i=1}^{n}\sum_{j=1}^{m} \left(\frac{R_{i,j}}{R_{max}} - \frac{U_i^T \cdot V_j}{||U_i|| \times ||V_j||}\right)^2 \tag{2}$$

Since $\frac{U_i^T \cdot V_j}{||U_i|| \times ||V_j||}$ is actually the cosine between vectors $U_i$ and $V_j$, we are actually looking for vectors in higher dimensions whose cosine values of angles between each pair are defined by $\frac{R_{i,j}}{R_{max}}$.

As in most real world data sets, $\frac{R_{i,j}}{R_{max}}$ follows power law distribution. To make the framework simpler, we assume the ratios follow Zipf Distribution. Namely, the frequency of occurrences of $\frac{R_{i,j}}{R_{max}}$ is proportional to their own values. This brings about a very interesting geometric property of the input data set : The majority values of $\frac{R_{i,j}}{R_{max}}$ are 1. This means to minimize L in Formula (2), most of $U_i$ and $V_j$ should be co-linear.

A natural question arises for us : What does the vector space of $U_i$ and $V_j$ look like ? Let's define the number of co-linear $U_i$ and $V_j$ pairs as M, then $\frac{M \times R_{i,j}}{R_{max}}$ of the vector pairs have cosine value $\frac{R_{i,j}}{R_{max}}$. We find it extremely difficult to come up with a visualization of such space by analytical methods without intervention of computers.

To examine the impact of the co-linearity property of the majority of $U_i$ and $V_j$, we propose an algorithm as follows : We define the loss function of the new algorithm (which we name Linear Factorization) below :

$$L = \sum_{i=1}^{n} \sum_{j=1}^{m} \left( \frac{R_{i,j}}{R_{max}} - \frac{U_i^T \cdot V_j}{||U_i|| \times ||V_j||} \right)^2$$

Subject to :

$$\begin{cases} \sum_{k=1}^{K} \alpha_k \cdot U_{1,k} = 0 \\ \cdots \\ \sum_{k=1}^{K} \alpha_k \cdot U_{n,k} = 0 \end{cases} \text{ and } \begin{cases} \sum_{k=1}^{K} \alpha_k \cdot V_{1,k} = 0 \\ \cdots \\ \sum_{k=1}^{K} \alpha_k \cdot V_{m,k} = 0 \end{cases} \quad (3)$$

We tested the algorithm on MovieLens 1 Million Dataset [19] (Fig. 1), and discovered that the technical accuracy of our proposed algorithm can achieve competitive results with other algorithms. By this observation, we safely draw the conclusion that the co-linearity property of user and item feature vector space plays a vital role in the technical accuracy of matrix factorization algorithms. We also find out that Linear Factorization is the most fair recommender system among the 9 algorithms. We will provide bibliographic information for the algorithms in this experiment in the Experiment Section.

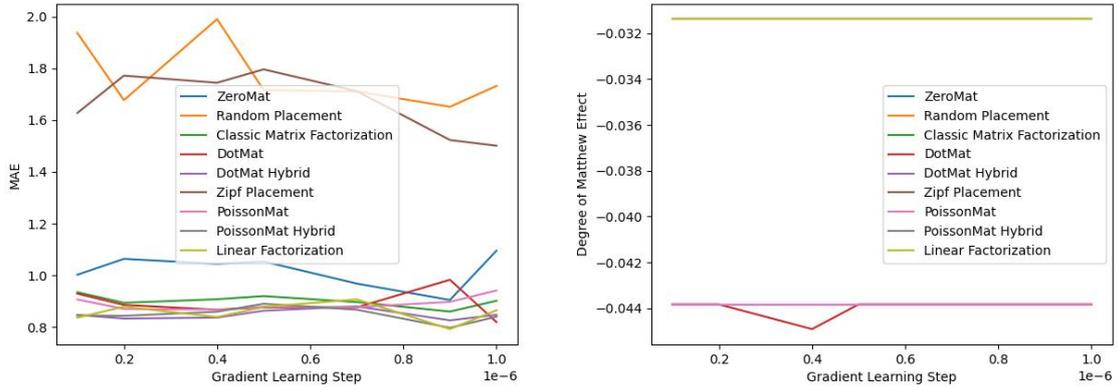

Fig. 1 Linear Factorization comparison experiments on MovieLens 1 Million Dataset (MAE and Degree of Matthew Effect)

However, due to the difficulty of designing a vector space functional that caters to all the geometric angle preserving vectors, we take a different route to solve the problem. Instead of considering $\frac{R_{i,j}}{R_{max}}$ as consine angles between vectors, we consider the value as the distance from a vector to the origin. By doing this, the input user item rating values (by Zipf Law) becomes equidistantly distributed sample points on co-centric circles.

The reason behind this geometric property is because the number of equidistantly distributed points on co-centric circles is proportional to the radius length. If we define the radii by $\frac{R_{i,j}}{R_{max}}$, the points of our newly designed geometry just become compliant with Zipf's Law.

To propose our new algorithm named ParaMat, we elevate our 2-D geometry into 3-D space by the following formula :

$$z = \frac{R_{i,j}}{R_{max}} \sqrt{x_k^2 + y_k^2} \tag{4}$$

We define the x and y as the products of user feature vector and item feature vector :

$$\begin{aligned} x_k &= U_i^T \cdot V_j \\ y_k &= W_i^T \cdot P_j \end{aligned} \tag{5}$$

The loss function for ParaMat is defined as follows :

$$L = \sum_{i=1}^{n} \sum_{j=1}^{m} \left( \frac{R_{i,j}}{R_{max}} - d\left( \left( x_k, y_k, \frac{R_{i,j}}{R_{max}} \sqrt{x_k^2 + y_k^2} \right), (0, 0, 0) \right) \right)^2 \tag{6}$$

We plug in the values of x and y by dot products of user feature vectors and item feature vectors, and substitute $\frac{R_{i,j}}{R_{max}}$ by dot products of user feature vectors and item feature vectors produced by the classic matrix factorization. We obtain the following formula :

$$L = \sum_{i=1}^{n} \sum_{j=1}^{m} \left( \frac{R_{i,j}}{R_{max}} - \sqrt{(U_i^T \cdot V_j)^2 + (W_i^T \cdot P_j)^2 + \left(\frac{R_{i,j}}{R_{max}}\right)^2 \left((U_i^T \cdot V_j)^2 + (W_i^T \cdot P_j)^2\right)} \right)^2 \quad (7)$$

We choose to optimize Formula (7) using Stochastic Gradient Descent (SGD) algorithm, after which we compute the unknown user item rating values using the following formula :

$$R_{i,j} = R_{max} \sqrt{(U_i^T \cdot V_j)^2 + (W_i^T \cdot P_j)^2 + \left(\frac{R_{i,j}}{R_{max}}\right)^2 \left((U_i^T \cdot V_j)^2 + (W_i^T \cdot P_j)^2\right)} \quad (8)$$

In the Experiment section, we show that ParaMat is the best algorithm when evaluated by fairness metric. After geometric analysis and derivation, we have obtained an effective solution for fairness problem in recommender systems.

## 4. EXPERIMENT

To evaluate the performance of ParaMat, we compare the algorithm with 8 other algorithms on both the MovieLens 1 Million Dataset [19] and LDOS-CoMoDa [20] dataset. Among the algorithms, Random Placement means recommend uniformly randomly; Zipf Placement means recommend according to Power Law distribution; ZeroMat [5] and DotMat [6] are introduced in the Related Work section; PoissonMat [21] is an algorithm to appear at an international conference in 2022.

Fig.2 illustrates the experimental results of comparison among 9 algorithms on the MovieLens 1 Million Dataset. ParaMat achieves the 5th place by technical accuracy metric MAE, but wins the 1st place by fairness metric Degree of Matthew Effect. By observation, ParaMat is a quite effective recommender system evaluated by fairness metric. Since the error curves are pretty cluttered, we also use Takens Embedding to visualize the curves in 2D (displayed in 3D grid for better visibility).

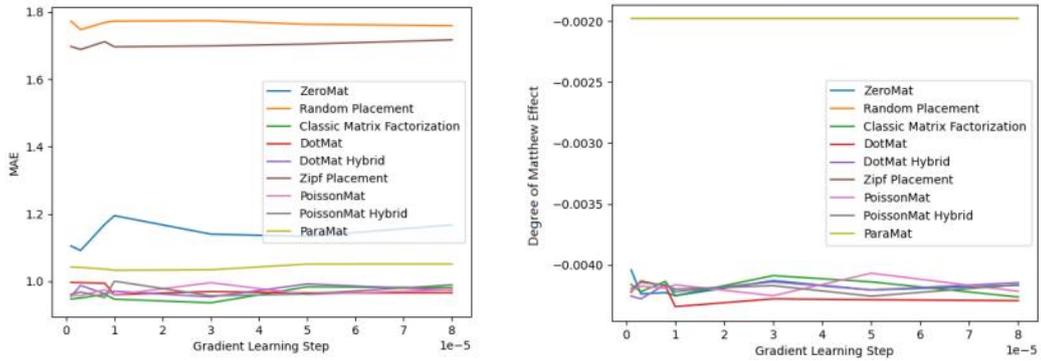

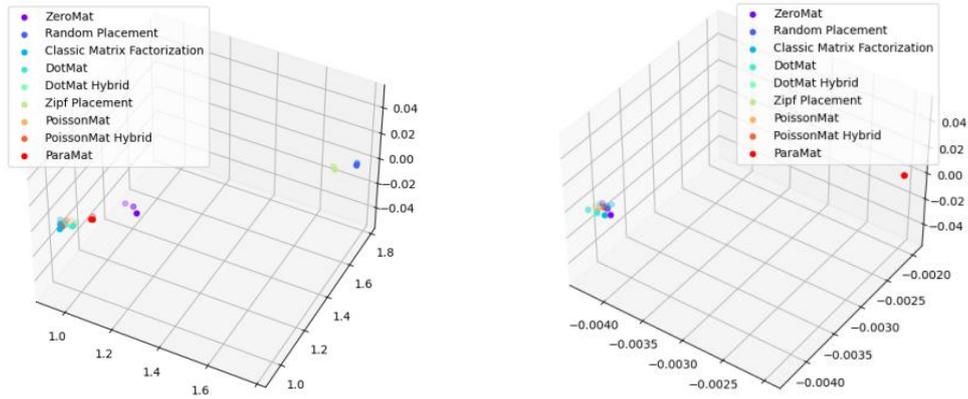

Fig. 2 ParaMat comparison on MovieLens 1 Million Dataset (MAE and Degree of Matthew Effect). The first row shows the MAE and DME curves of different algorithms. The bottom row visualizes the curves respectively using Takens Embedding.

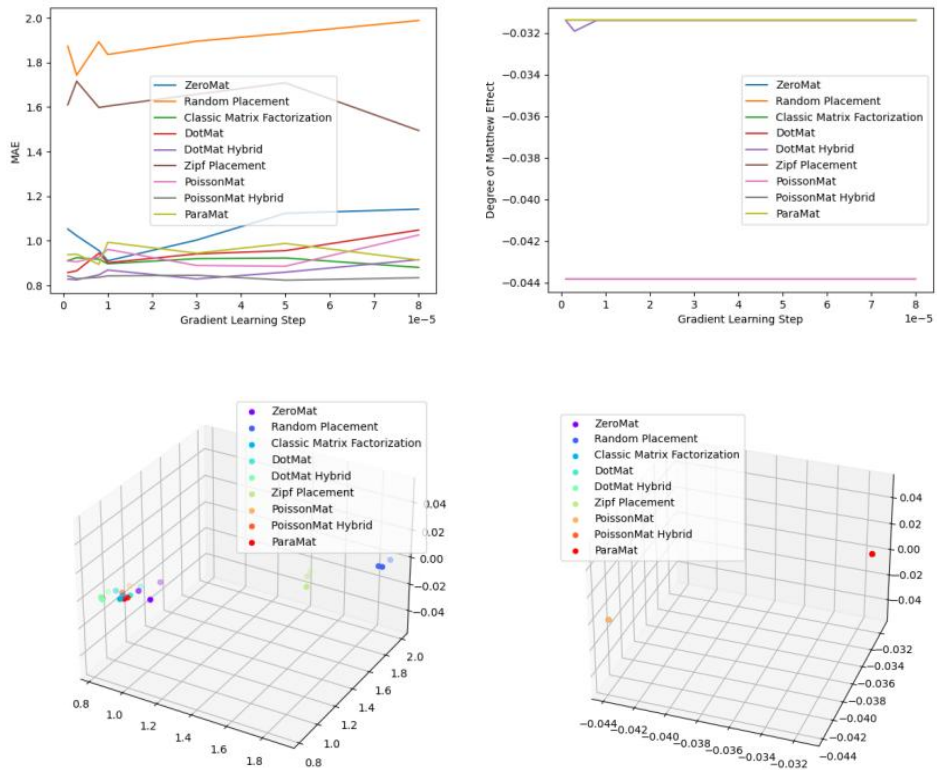

Fig. 3 ParaMat comparison on LDOS-CoMoDa Dataset (MAE and Degree of Matthew Effect). The first row shows the MAE and DME curves of different algorithms. The bottom row visualizes the curves respectively using Takens Embedding.

Fig.3 demonstrates the experimental results of comparison among 9 algorithms on the LDOS-CoMoDa Dataset. ParaMat achieves the 4[th] place by technical accuracy metric MAE, but once again wins the 1[st] place by fairness metric Degree of Matthew Effect. By observation, ParaMat is the most fair recommender system. The result is more clearly demonstrated in 2-D via Takens Embedding.

# 5. CONCLUSION

In this paper, we propose geometry-powered fair recommender system algorithms named Linear Factorization and ParaMat. Linear Factorization assumes the user feature vectors and item feature vectors of matrix factorization lay on the same hyperplane due to the observation that user/item feature vectors are mostly co-linear. ParaMat also relies on the geometric observation that user item rating values mostly lie on co-centric circles equidistantly.

Both Linear Factorization and ParaMat are not the best performing algorithm on technical accuracy metrics such as MAE, but they are the most fair algorithms among the 9 algorithms in our Experiment section. By analyzing the geometric space of the input data structure, we've come up with two effective fair recommendation algorithms.

In future work, we would like to explore the geometric space of input data structures to other classic algorithms such as learning to rank and factorization machines. We are also very interested in geometric interpretation of deep learning models.